# Digital Engineering Transformation as a Sociotechnical Challenge: Categorization of Barriers and Their Mapping to DoD's Policy Goals

Md Doulotuzzaman Xames | Taylan G. Topcu
Grado Department of Industrial and Systems Engineering, Virginia Tech, Blacksburg, VA 24061, USA

**Correspondence:** Md Doulotuzzaman Xames (xames@vt.edu)
**Funding:** None.
**Conflict of interest:** The authors declare they have no competing interests to disclose.

## Abstract

Digital Engineering (DE) transformation represents a paradigm shift in systems engineering (SE), which aims to synergistically integrate heterogeneous analytical models and digital artifacts into an authoritative source of truth for improved traceability and more efficient system lifecycle management. However, despite institutional support, many DE initiatives underperform or fail to realize their purported benefits. This paper posits that this is an outcome of poor understanding of social and technical barriers, and more specifically, how their interplay influences DE transformation efforts. To that end, we present a rich documentation of these challenges based on the literature and the sociotechnical systems theory, organized in dimensions of: people, processes, culture, goals, infrastructure, and technology. We then map these barriers against the U.S. Department of Defense's DE policy goals. Our findings reveal that technological investments alone are insufficient, and DE failures often stem from social factors such as workforce readiness, leadership support, and cultural alignment. Our mapping also reveals that the influence of sociotechnical barriers cascades across dimensions; thus, their impact on policy goals is difficult to trace, further complicating implementation efforts. These findings have direct implications for stakeholders who play a critical role in DE transformation initiatives. For instance, *managers* could use our mapping as a diagnostic tool to proactively identify risks and prioritize resource allocation; *policymakers* could complement their strategic mandates with sustained investments and long-term change management support; and finally, *engineers* could benefit from approaching DE as a more effective mode of collaboration, and not as a threat to their job security.

## 1 | Introduction

The increasing complexity of engineered systems [1,2], combined with repeated failures in complex system development program outcomes manifested in cost and schedule overruns [3–6], has challenged organizations to adapt by evolving their technologies, processes, and operations. To that end, digital transformation (DT) aims to leverage advanced tools and methodologies to revolutionize system lifecycle management by unlocking new efficiencies.

In the systems engineering (SE) community, particularly within the U.S. Department of Defense (DoD), DT is operationalized as "digital engineering" (DE). DE represents an integrated approach that relies on authoritative sources of system data and models, known as authoritative sources of truth (ASOT), to manage the system lifecycle from ideation through disposal [7]. ASOT provides a reference point for configuration-controlled models and data across the lifecycle, ensuring traceability, consistency, and reliability in information management [8]. Compared to traditional document-based SE, DE introduces a fundamental shift in *modus operandi*, using integrated digital models to design, analyze, and manage complex systems. This approach is expected to improve decision-making efficiency and help mitigate typical schedule delays and cost overruns [9,10].

While it is treated as a new trend in the SE community, DT is not unique to SE and has been adopted across various industries, including aerospace, manufacturing, healthcare, finance, retail, and agriculture [11,12]. However, these previous efforts have resulted in mixed results; as such, up to 90% of all DT initiatives fail to achieve their intended outcomes [13,14]. For SE organizations now grappling with similar transformations, these past experiences offer valuable lessons. Yet, transitioning toward a DE paradigm is easier said than done.

DE transformation is challenging for the SE Community, in part because SE organizations are sociotechnical systems (STSs) [15] with tightly interwoven social and technical components, as postulated in Conway's seminal *mirroring hypothesis* [16,17]. According to this hypothesis, the architecture of technical systems inevitably reflects the communication patterns and power structures of the organizations that design them [18]. Furthermore, this correspondence, or duality [19], is linked with the core SE function managing the interdependent tasks [20] (e.g., parallel development of subsystems), which persists because complex systems are only partially decomposable [21]. Consequently, successful DE transition requires attention not only to technology but also to organizational interfaces and decision-making structures that mirror and influence system architectures [18,22–24].

Despite these theoretical insights, much of the current funding and organizational efforts in DE transformation have disproportionately prioritized new tools and technologies [13], often neglecting the equally important social and organizational dimensions [14].

Unfortunately, this trend resembles the past experiences with model-based systems engineering (MBSE), where claimed benefits were frequently based on expectation rather than empirical evidence, and many adoption efforts faltered due to insufficient attention to sociotechnical factors [25,26]. Today, many SE organizations pursuing DE transformation also struggle to understand the multifaceted nature of the problem [27]. Consequently, despite growing institutional support and policy frameworks, especially within the DoD, DE efforts often encounter delays, fragmented adoption, or sometimes result in outright failures [9]. This paradox suggests a deeper misalignment between the promise of DE and the realities of organizational execution. These oversights point to a critical insight: successful DE implementation requires more than the deployment of advanced tools or collaboration platforms. Rather, it must address barriers that arise from the interdependencies between social, technical, and organizational domains.

To address these gaps, this study presents a systematic categorization of DE transformation barriers, grounded in sociotechnical systems theory, that could be useful for both theory-building and identifying effective change management strategies for SE organizations. To that end, this study investigates two research questions (RQs):

> **RQ1:** What are the key sociotechnical barriers that hinder the successful implementation of DE in organizations that develop and maintain complex engineered systems?
> **RQ2:** How do these sociotechnical barriers map onto the DoD's DE policy goals?

To answer these questions, we conducted a structured literature-based investigation to identify barriers to DE transformation. We then categorized these findings through qualitative coding by adopting a lens rooted in sociotechnical systems theory that is organized in six dimensions: people, technology, processes, culture, infrastructure, and goals. We then mapped each of these barriers to DoD's DE policy goals, to reveal critical misalignments between policy aspirations and implementation challenges that are manifested in practice. Our findings reveal the multifaceted and intertwined nature of these barriers, calling attention to the complexity of the DE transformation challenge that awaits many SE organizations. We contend that neglecting these interdependencies could negate the purported benefits of DE and introduce the risk of reproducing existing information silos or creating new coordination breakdowns. Related to this goal, this paper presents a rich discussion of these challenges, their implications for DoD's policy objectives, along with recommendations for practicing engineers, managers, policy-makers, and researchers to help facilitate a smoother DE transition.

## 2 | Background

### 2.1 | Digital Transformation and Digital Engineering

The evolution of DE can be understood from two distinct perspectives [28], as shown in Figure 1. The "digital evolution" perspective emphasizes how digital technologies are transforming SE through three stages: "digitization," "digitalization," and "digital engineering." From this lens, DE is a sociotechnical process focused on developing innovative business applications by integrating digitized data with digitalized systems [29]. This transformation is enabled by advances in information, computing, communication, and connectivity technologies, which collectively drive fundamental changes in enterprise operations [30]. Characterized by iterative processes and rapid adaptation, DE requires participatory approaches [31], encouraging a shift from management-driven to data-driven decision-making. In doing so, it fosters interdisciplinary collaboration, agility, experimentation, and helps counter risk-averse practices [32].

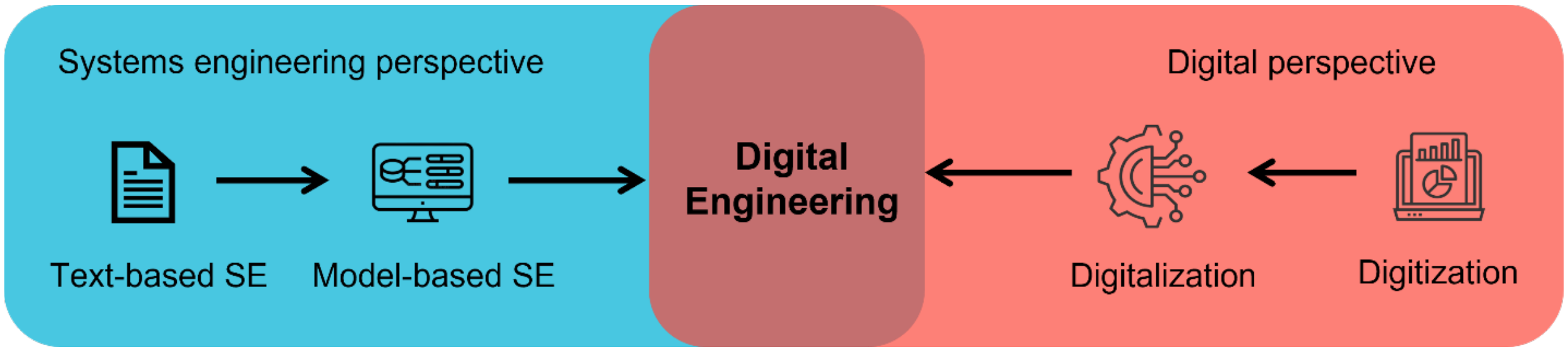

**Figure 1** | The development of the DE concept (adapted from [28]).

From the SE perspective, DE extends MBSE by integrating digital twins and digital threads [33], enabling continuous system assessment, enhanced data exchange, and improved collaboration across lifecycle stages. Several scholars position DE as the next evolutionary step beyond MBSE, emphasizing model-centric integration across design, development, testing, and sustainment [28,34]. In this sense, DE represents both a continuation of MBSE and a specialized application of DT within SE – one that aligns with the DoD's definition of DE as an authoritative, model-based approach to lifecycle management [7,8]. This perspective underscores that DE is not only about tools but also about establishing a resilient digital infrastructure capable of sustaining complex system development over decades.

Unlike DT in other sectors (e.g., healthcare, retail), DE in SE entails unique challenges. First, the inherent complexity of engineered systems necessitates collaboration of extremely large design teams that are distributed across numerous organizations, often for extended periods of time [35]. The interdisciplinary knowledge required to develop modern engineered systems exceeds the information processing capacity of any engineer, and necessitates the collaboration of subject matter experts from various domains to make informed trade-offs [36,37]. This dependence, in turn, creates a need for seamless integration of diverse analytical tools and languages [21]. Second, complex system development takes extended periods of time, which could be in the order of decades in some cases [38], creating an information continuity challenge across the lifecycle. Post-deployment, such systems typically operate for decades and undergo numerous post-production design changes [39] – usually driven by evolving mission needs, emerging new capabilities, and obsolescence [34,40]. Finally, DE in SE aims to leverage digital twins and digital threads to enable integrated lifecycle management, ensuring continuous synchronization across the design, development, testing, and maintenance phases of complex systems [41,42]. This necessitates increasingly more diverse multimodal data to be aggregated and maintained.

## 2.2 | The High Failure Rates of Digital Transformation Efforts Across Industries

Despite significant investments and widespread interest across industries, up to 90% of DT transformation initiatives fail to achieve their intended outcomes [13,14]. These failures are not merely isolated cases but rather indicative of systemic issues that persist across sectors. In the context of SE, this statistic should serve as a cautionary signal, especially for organizations pursuing complex, large-scale digital initiatives. Following the DoD's 2018 Digital Engineering Strategy [7], numerous DE initiatives have been launched across various DoD agencies and among their contractors, signalling a concerted effort to shift toward a more integrated, model-based approach to SE. However, many organizations continue to struggle with these transformation efforts. These failures are documented to result in dire consequences, including increased costs, extended timelines, poor design integration, elevated risk of mission failure, and a loss of competitive advantage in defense innovation [32], [34]. To that end, understanding and addressing these persistent failure modes is essential for unlocking the true potential of DE transformation.

Related to this goal, the SE community's prior struggles with MBSE adoption provide a useful parallel. Despite significant interest and investment, MBSE adoption, usage, and realized return on investment (ROI) have often fallen short of expectations. Studies found that most claimed benefits of MBSE were based on perceived rather than measured evidence [25,26], highlighting a persistent gap between aspiration and reality. This pattern underscores the risk that DE transformation may follow a similar trajectory if its sociotechnical complexities are not adequately addressed.

## 2.3 | The Need for a Sociotechnical Perspective

The STS theory originated in the 1950s at the Tavistock Institute, aimed at democratizing work processes, enhancing job satisfaction, and infusing humanistic ideals into industrial work practices [45,46]. Initially focused on manual labor, the STS concept has evolved to accommodate increasing technological complexity, including automation, computing, artificial intelligence, and social media [47]. As illustrated in Figure 2, STSs consist of two interdependent subsystems: a social subsystem, which includes individuals with their knowledge, skills, attitudes, values, and needs within the organizational context, and a technical subsystem, comprising the tools, mechanisms, and techniques employed to perform tasks [48]. Achieving coherence between these subsystems is essential in DE to optimize outcomes like productivity, system efficiency, and workforce engagement [46].

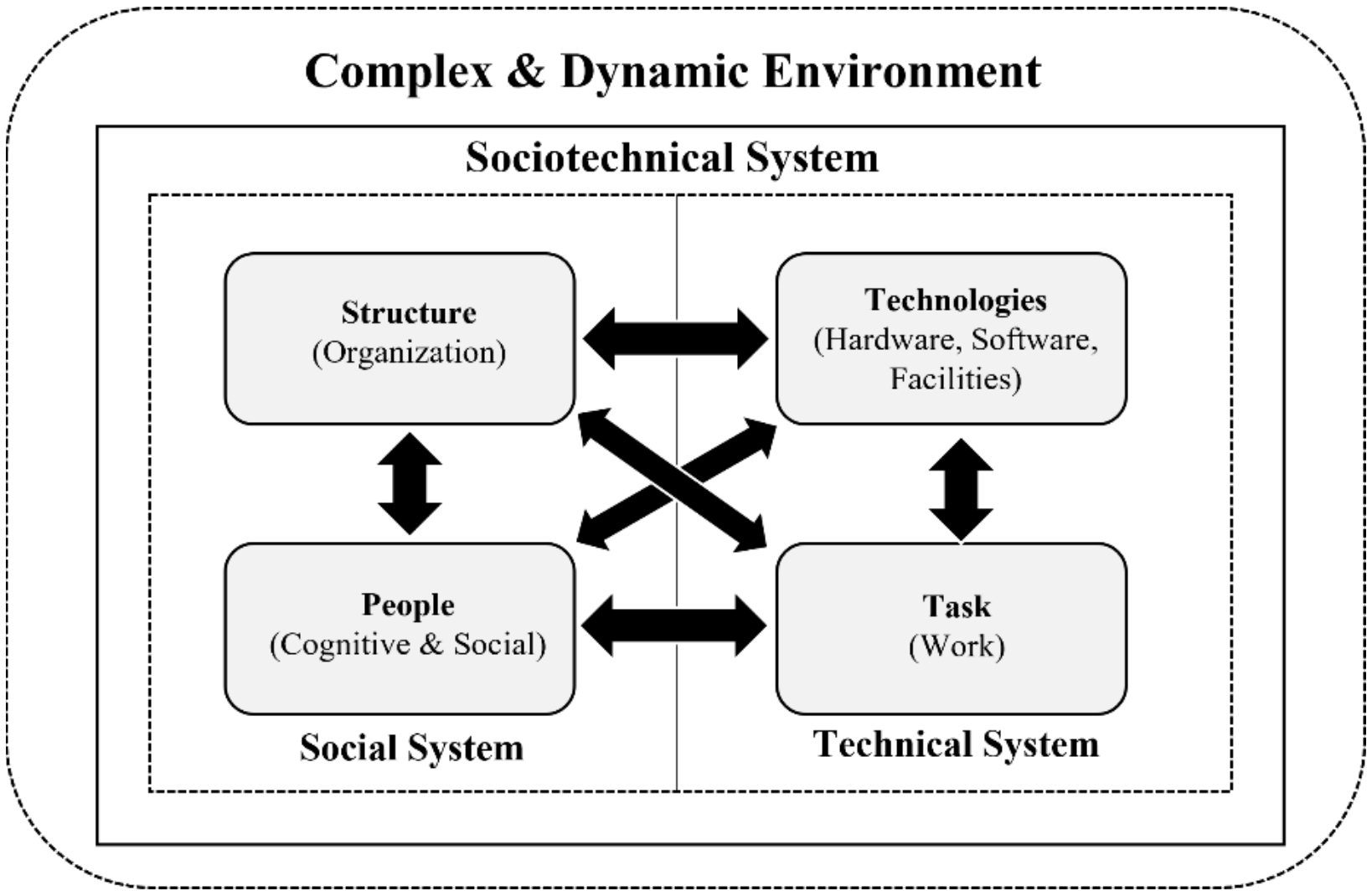

**Figure 2** | The interacting components within a sociotechnical system (adapted from [49]).

Over the past decades, several STS frameworks have been proposed to operationalize these interdependencies and guide the analysis of change in complex systems. One of the earliest was Leavitt's Diamond [50], which conceptualized organizations with four interacting elements: people, tasks, structure, and technology. Leavitt emphasized that a change in one element (e.g., technology) necessitates changes in others to maintain system balance. Similarly, Emery and Trist [45] framed organizations as open systems embedded in a broader environment and highlighted the importance of jointly optimizing both social and technical subsystems within this broader ecological context.

Building on these foundations, Bostrom and Heinen [49] formalized the concept of sociotechnical design in information systems, proposing two core subsystems (technical and social) and recommending participatory design practices to align them effectively. Later, Morton [51] introduced a more managerial-oriented perspective with five elements: strategy, structure, people, processes, and technology. His framework is particularly relevant for large-scale organizational transformation, as it integrates strategic intent with operational realities, which reflects Conway's Law by emphasizing that organizational structures and communication patterns shape the design and implementation of technical systems [19]. More recently, Davis et al. [52] proposed a framework that classifies sociotechnical systems into six dimensions: people, technology, processes, culture, infrastructure, and goals. While there are nuances, the common denominator across these perspectives is the interplay between the social and technical spheres, and the need to cater to them concurrently. This view is in line with Conway's Law and plays a significant role in the SE community [17–19].

## 2.4 | Research Gap

DE is increasingly recognized as a mechanism for value creation, where organizations reimagine products and services as digitally enabled assets within a sociotechnical environment that integrates technology into existing structures [53]. However, DE efforts often fall short not because of technical limitations, but due to the misalignment between technological change and existing work structures, cultural norms, and organizational processes [54]. As such, DE transcends the mere adoption of new tools; it requires a fundamental restructuring of how people, technologies, and institutions interact. Thus, viewing DE through an STS lens is therefore critical for uncovering the tensions between social and technical spheres that are essential for successful transformation [55].

Nevertheless, existing research on DE has paid insufficient attention to its sociotechnical characteristics. Currently, most DE transformation efforts continue to emphasize tool development, platform integration, or model-centric workflows, often at the expense of human and organizational considerations such as workforce readiness, leadership support, cultural resistance, and strategic goal alignment. As DE becomes increasingly central to how complex systems are conceived, developed, and sustained, its implementation challenges demand closer scrutiny.

In addition, although the DoD has articulated clear policy goals to advance DE, ranging from model integration to cultural change, there is a lack of research that connects DE transformation barriers to these policy objectives. Without this linkage, it becomes difficult for practitioners and policymakers to assess where gaps exist, identify the secondary mechanisms that need to be addressed concurrently, and prioritize interventions. Hence, this disconnect between operational challenges and strategic goals limits the effectiveness of DE implementation efforts. This paper addresses this gap at its core by providing a theory-informed approach that categorizes sociotechnical barriers to DE transformation and then maps these barriers to DoD's stated policy goals, in pursuit of enabling more targeted and systemic interventions.

# 3 | Materials and Methods

To address our research questions, we employed a four-step approach, as illustrated in Figure 3. First, we conducted a rapid literature review to identify barriers to DE transformation. Next, we synthesized and consolidated these barriers through inductive coding. We then categorized the barriers using a sociotechnical framework comprising six interrelated dimensions. Finally, we mapped the categorized barriers to the DoD's DE policy goals to assess alignment and reveal implementation challenges. Below, we elaborate on each of the four steps of our research methodology.

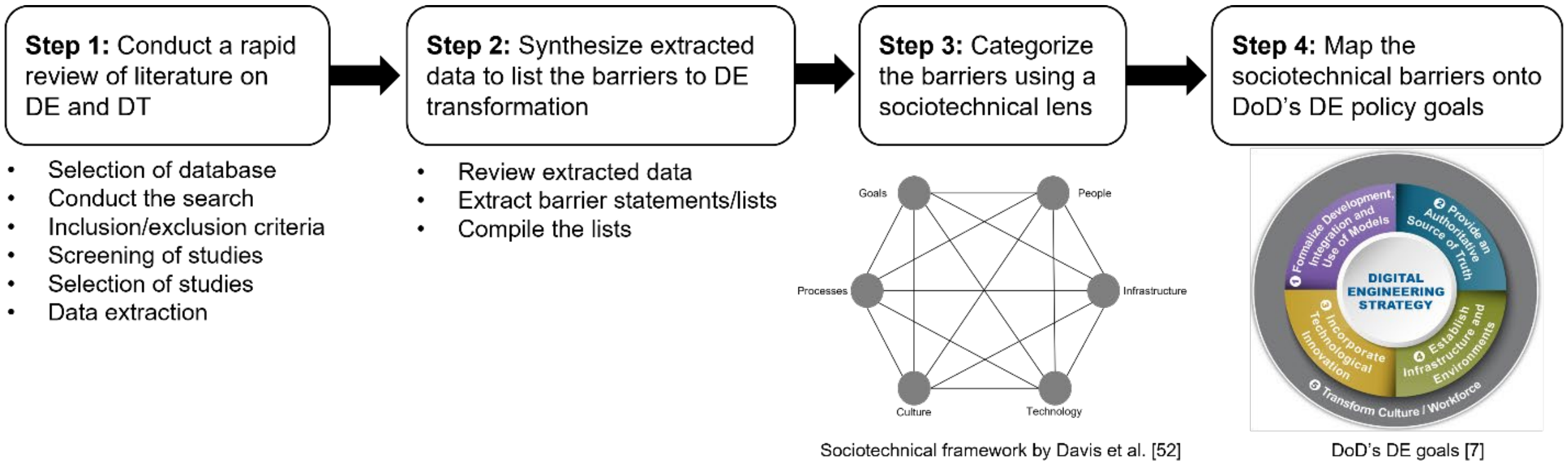

**Figure 3** | The four-step research methodology used in this study.

In the *first* step, we conducted a rapid review of the literature on DE and DT, following the approach outlined by Khangura et al. [56]. We utilized *Google Scholar* as our primary search database to conduct a literature search, capturing relevant publications published from 2016 to May 2025. *Google Scholar* was chosen for its broad coverage across peer-reviewed journals, conference proceedings, theses, technical reports, and grey literature [57], making it particularly suitable for interdisciplinary topics like DE. Our search strings included combinations of keywords related to "digital transformation" and "digital engineering," combined with barrier-related keywords such as "barriers", "challenges", and "obstacles". In the initial search, sorted by relevance, we considered all publications drawn from the first 10 pages of results per keyword combination. We then screened titles and abstracts to identify relevant studies. After applying inclusion criteria such as studies that discussed barriers to DE or DT, we retained around 31 studies for full-text review. From these, we extracted 49 unique barriers that are pertinent to DE transformation. Rather than confining our observations solely to SE organizations, we broadened our scope to encompass various application areas, allowing us to gather diverse insights and perspectives. This approach enabled us to understand the barriers associated with DTs across multiple industries, insights that are equally relevant in the context of DE transformation.

In the *second* step, we synthesized the retrieved literature and then extracted barriers pertinent to DE through an inductive coding. This approach is more appropriate in such cases because it allows researchers to remain open to identifying the most relevant factors, rather than being constrained by predefined variables [58], and it has been shown to be effective for uncovering patterns from raw text data [59]. We then compiled the barriers list to remove redundancies and finalize our unique list of barriers, ultimately resulting in 49 barriers.

The *third* step involved applying the STS framework proposed by Davis et al. [52] (shown in Figure 4), which conceptualizes six interdependent dimensions within complex organizational systems:

- **Culture:** Refers to the shared beliefs, values, norms, and assumptions that influence behavior and decision-making within an organization.
- **Goals:** Captures the strategic intent, vision, and objectives that guide organizational behavior.
- **Infrastructure:** Includes the foundational physical and digital systems, resources, and logistical arrangements that support operational functioning.
- **People:** This dimension refers to the human actors within an organization, including their skills, knowledge, roles, attitudes, motivations, and interpersonal dynamics.
- **Processes:** Defined as the formal and informal workflows, procedures, and task sequences that structure how work gets done.
- **Technology:** Encompasses the technical artifacts, tools, systems, and platforms used to perform tasks and enable communication.

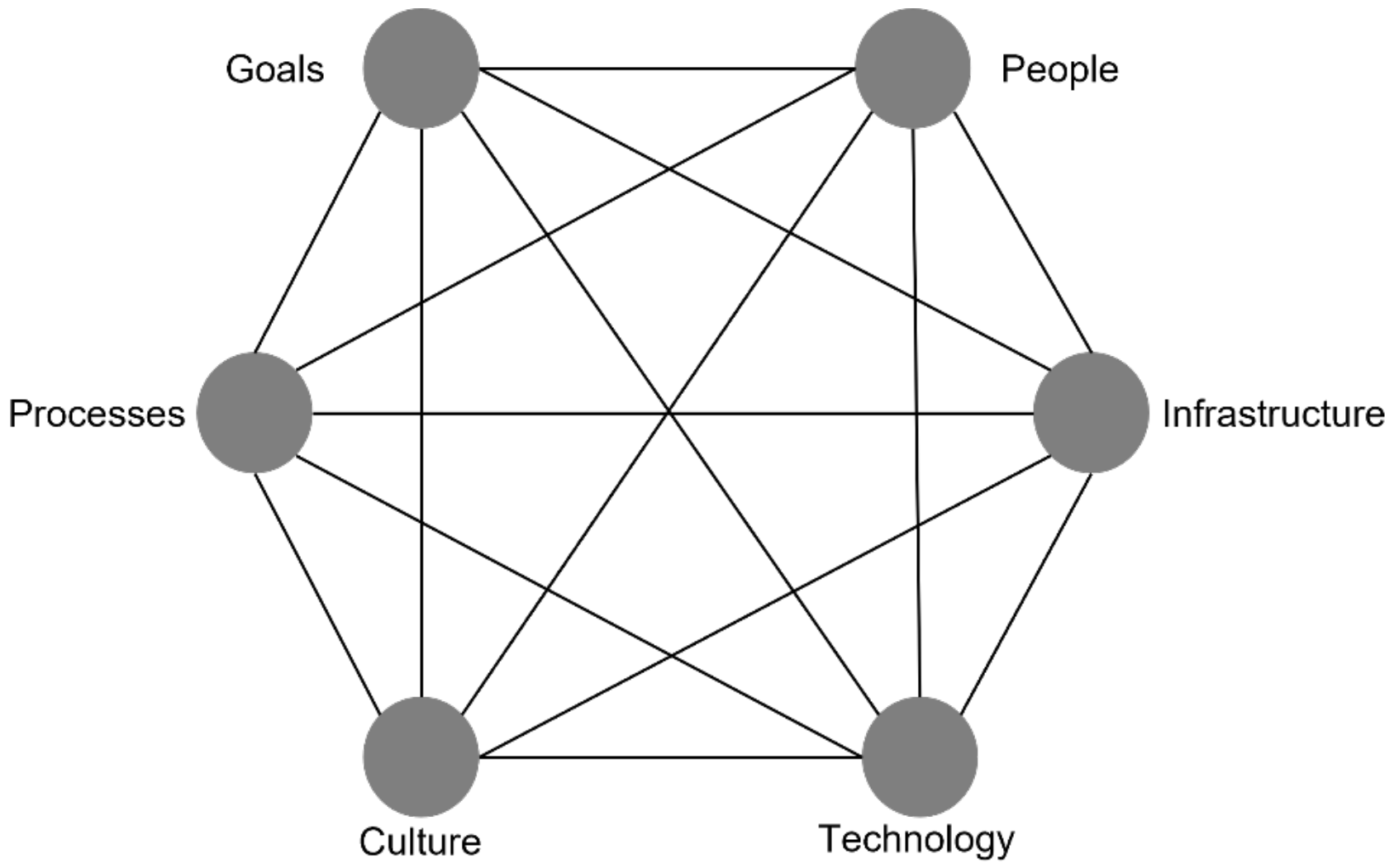

**Figure 4** | Sociotechnical framework within a complex system proposed by Davis [52].

We selected this framework because it provides a holistic and contemporary lens that captures the complex interdependencies between social and technical elements within organizations. Unlike earlier STS frameworks, it explicitly incorporates strategy (goals) and infrastructure, two dimensions that are critically relevant to DE transformation in large, technologically intensive, and mission-driven systems. By emphasizing interdependencies across people, technology, processes, culture, infrastructure, and goals, the framework offers a nuanced understanding of barriers that aligns with the cross-cutting nature of DE challenges. Moreover, its use in recent applied ergonomics and systems thinking literature underscores both its theoretical grounding and its practical relevance for analyzing organizational change in high-tech contexts.

In this step, the 49 identified barriers were categorized into the six sociotechnical dimensions proposed by Davis et al. [52]. Most barriers were assigned directly to the dimension they most closely resembled. In cases where a barrier could be argued to correspond to more than one dimension, it was grouped with the dimension to which it had the closest alignment.

In the *fourth* and final step, we extracted DoD's DE policy goals from the DoD's Digital Engineering Strategy guide and then mapped the sociotechnical barriers identified in Step 3 onto these policy goals. Each barrier was evaluated for its potential impact on achieving each DE goal, and we assigned it to the goal(s) where it posed a significant implementation challenge. As shown in Figure 5, the DoD's DE strategy outlines five transformation objectives: (i) formalize the development, integration, and use of models; (ii) provide an ASOT; (iii) incorporate technological innovation; (iv) establish infrastructure and environments; and (v) transform the culture and workforce. Each of these goals represents a critical pillar in the DoD's vision for DT, yet their successful implementation depends on the resolution of multiple sociotechnical challenges. Therefore, our mapping provides insights into the intricate relationships between sociotechnical challenges and each of the DoD's DE transformation objectives. By understanding these connections, the DoD, its various agencies, and contractors in the industry can better prioritize resources, make their strategic decisions, and more efficiently implement DE initiatives.

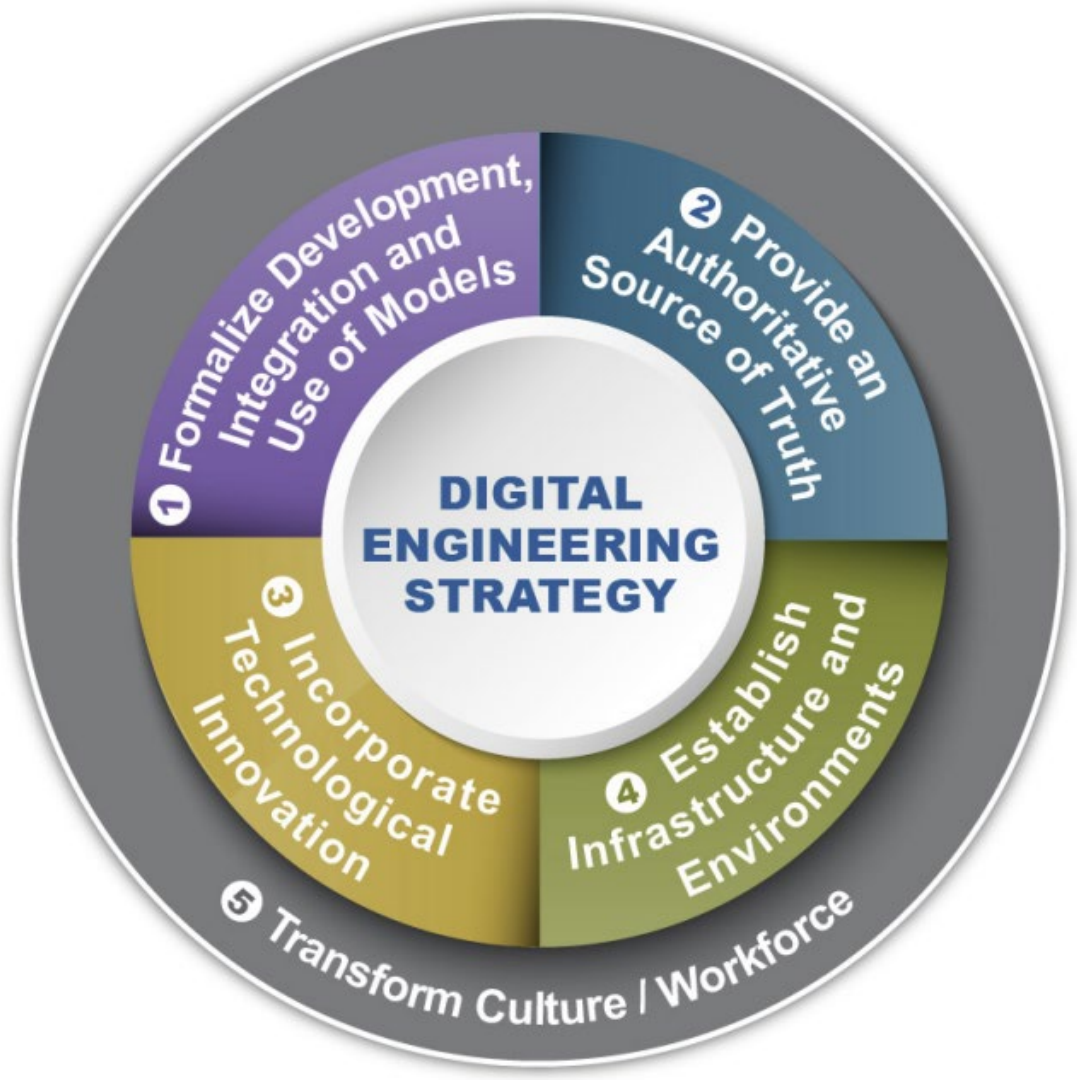

**Figure 5** | Department of Defense's DE goals [7].

## 4 | Results

Through our rapid review and inductive coding, we identified 49 unique barriers to DE transformation. As illustrated in Figure 6, the distribution of these barriers across the six sociotechnical dimensions reveals that culture-related barriers are the most commonly discussed ($n = 11$), closely followed by people-related barriers ($n = 10$). This frequency highlights the significant role of organizational norms, leadership behavior, resistance to change, and workforce readiness in obstructing DE transformation. Technology, process, and goal-related barriers each account for eight challenges, indicating their common and relatively balanced presence. Infrastructure-related barriers are the least represented ($n = 4$), suggesting that while foundational, infrastructural issues may be less frequently cited than social or organizational impediments. Overall, the findings underscore the dominance of human and cultural factors in shaping the success or failure of DE initiatives, reinforcing the need for a sociotechnical approach that goes beyond tool deployment and emphasizes organizational change management. Next, we discuss how these findings relate to our RQs.

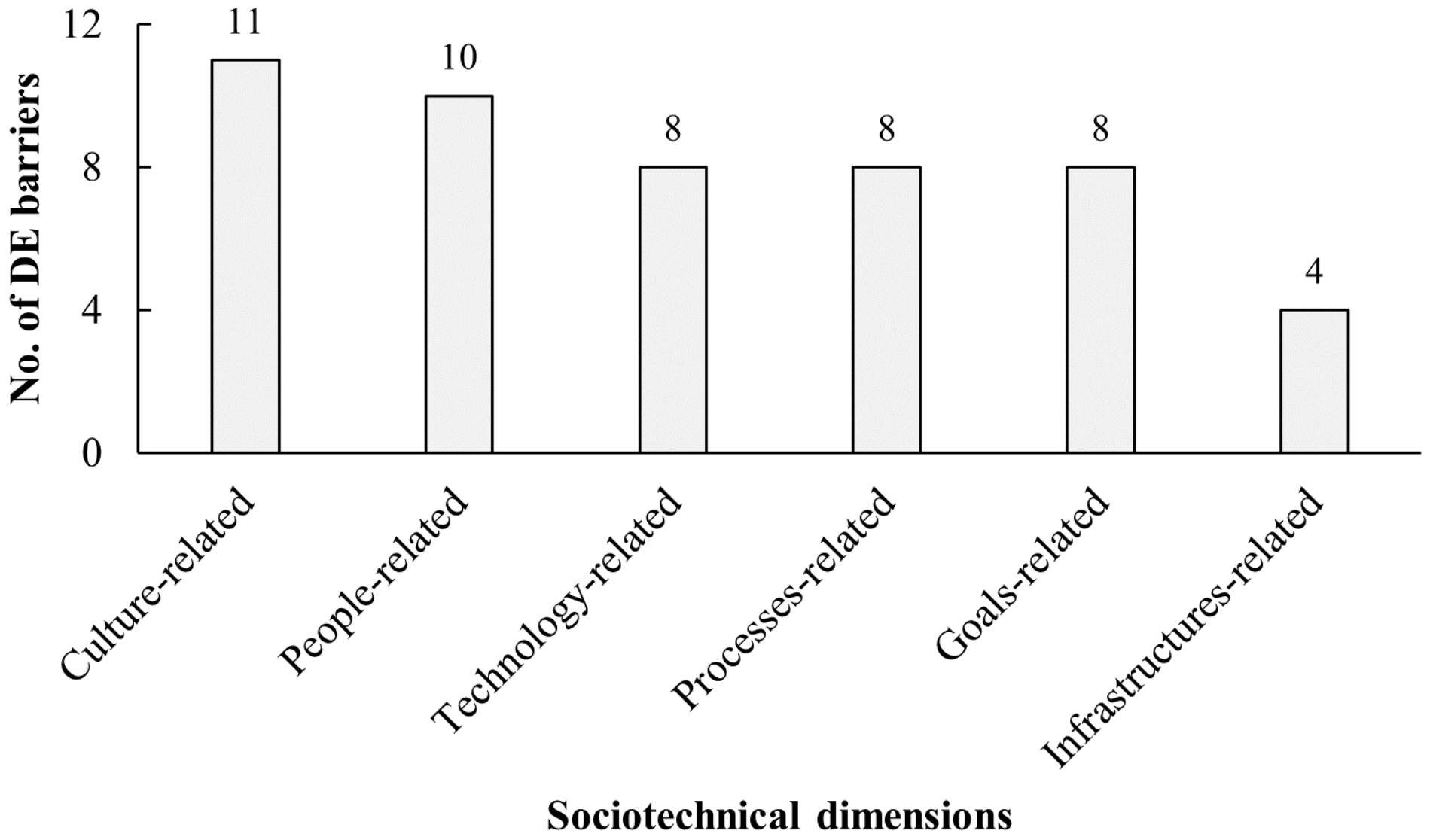

**Figure 6** | Histogram showing the number of DE transformation barriers identified across six sociotechnical dimensions ($n$=49).

## 4.1 | Findings for RQ1: Categorization of Sociotechnical Barriers to DE Transformation

In Table 1, we present our categorization of the 49 barriers into six sociotechnical dimensions, emphasizing their interrelated nature and the need for addressing these complex interdependencies to increase the probability of success of DE transformation efforts. For example, the effective implementation of technology depends on the readiness and skills of the workforce, while organizational culture can either support or resist change. Additionally, aligning goals with technological capabilities and infrastructure is essential for driving effective processes. Each of the six dimensions comprises a range of distinct and nuanced barriers. For instance, the people category includes ten differentiated factors, such as workforce capacity, digital literacy gaps, resistance to change, and communication breakdowns. Similarly, barriers within the culture dimension span leadership mindset, risk tolerance, and organizational silos. This breakdown reveals the granular nature of challenges even within a single dimension, highlighting the importance of distinguishing between specific subthemes when diagnosing barriers to DE transformation.

**Table 1** | Sociotechnical barriers to digital engineering transformation.

| **Sociotechnical dimensions** | **Barriers to digital engineering transformation** | |
|---|---|---|
| *Culture* | ▪ Resistance to change [60–64]<br>▪ Lack of leadership for change [65,66]<br>▪ Organizational inertia [29,64]<br>▪ Risk aversion attitude [60,61,67]<br>▪ Lack of commitment [29,55]<br>▪ Unwillingness for upfront investment [68,69] | ▪ Silo thinking [64,65]<br>▪ Bureaucratic culture [62,65,70]<br>▪ Lacking customer pool [29,70]<br>▪ Lack of managerial support [61,62]<br>▪ Lack of trust in external stakeholders [69]<br>▪ Lack of innovative spirit [70] |
| *Goals* | ▪ Inadequate IT infrastructure [65,71,72]<br>▪ Lack of IT support [65]<br>▪ Lack of financial resources [29,60] | ▪ Lack of technological readiness [64]<br>▪ Lack of time for transformation activities [60,73] |
| *Infrastructures* | ▪ Lack of clear vision/strategy [29,60,65]<br>▪ Unstable or misaligned decision-making [70]<br>▪ Lack of institutional policy [65]<br>▪ Lack of change management [69,70] | ▪ Uncertainty about ROI [29,65,69]<br>▪ Focus on reducing costs [61]<br>▪ Focus on maximizing profit [61] |
| *People* | ▪ Lack of digital literacy [60,61,65]<br>▪ Lack of communication and coordination [61,62]<br>▪ Lack of personnel [62]<br>▪ Insufficient training [61,68,73]<br>▪ Hiring and retaining a qualified workforce [74] | ▪ Lack of process knowledge [60,62]<br>▪ Fear of job loss [60]<br>▪ Fear of loss of control [60,68]<br>▪ Fear of using technology [62]<br>▪ Obtaining workforce clearance [74] |
| *Processes* | ▪ Lack of standards [60,75]<br>▪ Data quality and management issues [68,71]<br>▪ Process complexity [61,62,70]<br>▪ Poor estimation of total costs [74] | ▪ Restrictive laws limiting process flexibility [60,76]<br>▪ Process interoperability issues [61,68]<br>▪ Legal compliance issues [65,75]<br>▪ Long time period between implementation and results [77] |
| *Technology* | ▪ Security and privacy issues [65,68]<br>▪ Usability issues [61]<br>▪ Dependence on other technologies [60]<br>▪ Resolving intellectual property issues [74] | ▪ Technological disruptions [65]<br>▪ Isolated systems [65,70]<br>▪ Lack of data ownership [78] |

Below, we provide a succinct discussion of the barriers with respect to each dimension.

Rather than implying any prioritization, we are simply following the alphabetical order in our reference categorization scheme in Table 1.

#### 4.1.1 | Culture-related Barriers

Cultural factors play a pivotal role in shaping individuals' attitudes and behaviors within organizations, thus greatly influencing the success of DE initiatives. Resistance to change is a pervasive barrier [68], as employees may cling to familiar routines and processes [64], fearing the uncertainty that comes with innovation. Cultural resistance is often sustained not just by individual attitudes but also by deeply embedded organizational norms and leadership behaviors [61]. For instance, hierarchical decision-making environments can discourage bottom-up innovation and create bottlenecks for experimentation with digital tools [62]. Similarly, organizational inertia and risk-aversion can restrain innovation [60], perpetuating the status quo and impeding progress. Additionally, the tendency for silo thinking [64] – where departments or teams work in isolation with limited communication – and lack of collaboration can inhibit the cross-functional cooperation that is necessary for successful DE transformation [70]. Related to these points, a lack of leadership support for change [65,66] is documented to reinforce siloed thinking, while insufficient acknowledgment or incentives for cross-functional collaboration can discourage teams from sharing knowledge and adopting best practices across the organization [64,65]. Furthermore, in risk-averse or bureaucratic cultures, which are arguably typical in many government organizations, even well-supported DE initiatives may stall if mid-level managers perceive them as conflicting with performance metrics or established workflows.

#### 4.1.2 | Goals-related Barriers

The goals dimension encompasses the strategic objectives and outcomes organizations aim to achieve through DE transformation. A lack of clear vision and strategy can derail initiatives, leaving stakeholders confused about the purpose and direction of DE transformation efforts [65,69,70]. Without a roadmap for success, organizations risk meandering aimlessly or pursuing disjointed initiatives that fail to deliver meaningful results. Additionally, uncertainty about the ROI can deter stakeholders from committing to change [79], especially if they perceive DE transformation as an endeavor with uncertain benefits [65,69]. The barriers to DE transformation also include a narrow focus on cost reduction, and particularly in the case of industry contractors, a strong emphasis on maximizing profit, which can result in underinvestment in innovative technologies, potentially obstructing the adoption of DE initiatives [61].

Strategic misalignment often manifests itself as fragmented or overly narrow goals, with each organizational unit pursuing its own objectives under loosely or sometimes poorly coordinated strategies [65]. This results in duplicated efforts, inconsistent tool adoption, and miscommunication around priorities. In some cases, goals may focus disproportionately on achieving technical milestones, such as model integration or cost savings, while neglecting softer but equally critical outcomes like cross-functional collaboration, learning, or long-term capability building. Unstable or misaligned decision-making processes further amplify these issues [70], particularly when leadership shifts disrupt continuity in vision or accountability.

#### 4.1.3 | Infrastructures-related Barriers

The infrastructure dimension encompasses the physical and technological resources necessary to support DE transformation initiatives. Among these, inadequate IT infrastructure and technological readiness pose significant barriers, limiting organizations' ability to implement and scale digital solutions effectively [65]. Without the necessary hardware, software, and technical infrastructure capabilities, organizations may struggle to modernize their operations and compete in a digital-first landscape [71]. Insufficient IT support amplifies these challenges by further frustrating users, who are already struggling to effectively leverage the technology, creating a compounding effect on overall system efficiency [80].

Moreover, a lack of financial resources and time constraints, such as pressure for immediate results, further exacerbate these challenges, impeding investments in infrastructure upgrades and training programs [29,60]. Time constraints also limit organizations' capacity to engage in transformation activities, such as piloting infrastructure upgrades or planning phased adoption strategies [60,73]. When day-to-day operational demands dominate attention, as is typical in many complex system development programs, there is little bandwidth for the incremental adjustments needed to modernize infrastructure.

#### 4.1.4 | People-related Barriers

The human element plays an integral role in either facilitating or hindering DE transformation. One significant barrier is the lack of digital literacy and process knowledge among employees [62,65]. Without a fundamental understanding of new digital tools at their disposal and the organizational processes they aim to enhance, individuals may feel overwhelmed, leading to resistance to change. This lack of knowledge can impede their ability to embrace new technologies and adapt their workflows accordingly. Moreover, the fear of job loss exacerbates this reluctance [60]. Employees may perceive the transformation as a threat to their livelihoods, fearing that automation and digitization will render their skills obsolete [62].

Beyond these, systemic workforce challenges also contribute to persistent people-related barriers. Organizations may often face difficulty hiring and retaining individuals with specialized DE expertise [74], particularly in competitive labor markets where industry demand for such talent is rapidly increasing. Specifically for the SE community, clearance requirements that are typical in the aerospace and defense industry further restrict the eligible talent pool [74]. Additionally, limited internal capacity for training and upskilling exacerbates disparities in digital fluency, which hinders the integration of cross-disciplinary teams needed for DE success [61]. These human capital constraints are further intensified by communication breakdowns between technical and non-technical staff, which result in misaligned expectations and poor collaboration across lifecycle phases.

#### 4.1.5 | Processes-related Barriers

In DE transformation, there is a need to streamline and optimize processes for greater organizational efficiency and agility. However, several process-related barriers hinder this endeavor, including the lack of standards and interoperability issues [29]. Incompatible systems and disparate data formats impede seamless integration of the SE process, hampering data exchange and workflow automation, and complicating requirements traceability – critical for maintaining alignment between system requirements and technical development across lifecycle stages [68,71,75]. Moreover, data quality and management issues undermine the reliability of DE processes [61], leading to suboptimal decision-making and performance.

Additionally, DE transformation often introduces new layers of process complexity before realizing its promised efficiencies. For example, legal and regulatory constraints, such as data sovereignty laws or export controls, can limit the ability to standardize digital workflows across organizational or national boundaries [60,76]. Moreover, the long delay between DE implementation and observable performance outcomes makes it difficult for organizations to justify investments, particularly when budgets are structured around short-term deliverables [77]. This challenge is often amplified in contexts where leadership rotates frequently, as is common in government SE organizations, resulting in limited appetite for pursuing long-term structural changes. Furthermore, organizations often struggle with accurately estimating the full range of costs associated with DE transformation [74]. These “hidden” lifecycle costs, indirect dependencies, and unanticipated process changes can lead to underestimation, making it difficult to justify investments and to plan for allocating sufficient resources.

#### 4.1.6 | Technology-related Barriers

The technology dimension presents its own set of challenges in DE transformation, ranging from security concerns to usability issues [61,68]. Security and privacy issues loom large in an era of increasing cyber threats and strict data protection regulations [68]. Without robust security measures in place, organizations risk exposing sensitive information, which may erode trust among stakeholders [61]. Additionally, usability issues can impede adoption, particularly when employees struggle to navigate complex interfaces or encounter system glitches and bugs [68]. Technology-related barriers are not solely rooted in the tools themselves but also in the fragmented digital ecosystems in which they operate and the interfaces they have to form with legacy tools and processes. Success in DE transformation requires seamless integration; however, organizations often rely on legacy platforms that lack compatibility with emerging tools, necessitating extensive customization or middleware solutions to enable integration [60]. Lack of clarity regarding data ownership further impedes interoperability, especially when multiple contractors, vendors, or organizational units share responsibility for the engineering effort [78].

Moreover, unresolved intellectual property (IP) concerns may create friction during the sharing of analytical models or the development of collaborative tools, which may slow down the SE process, particularly given its iterative nature [74]. These challenges underscore the importance of not only investing in cutting-edge tools but also aligning digital infrastructure with governance protocols and stakeholder agreements that enable seamless, secure, and collaborative technology creation and use across organizational boundaries.

Taken together, these findings highlight the multifaceted and interdependent nature of barriers to DE transformation. Having established this categorization, we now turn to examine how these barriers relate to the DoD's DE policy goals.

### 4.2 | Findings for RQ2: Mapping of Sociotechnical Barriers onto DoD's DE Policy Goals

The DoD has identified five DE policy goals essential for successful transformation [7]. However, as discussed in the previous section, nuanced sociotechnical barriers impede progress toward these objectives. We contend that mapping these barriers to DE policy goals could be vital for developing a more targeted approach to implement DE practices and address specific challenges associated with them. For example, by understanding how issues such as digital literacy gaps or resistance to change obstruct particular DE policy goals, organizations can allocate resources more efficiently, directing interventions where they are most needed. Our mapping is summarized in Table 2.

**Table 2** | Mapping of sociotechnical barriers onto DE goals.

| DE goals | *Sociotechnical dimensions* | | | | | |
|---|---|---|---|---|---|---|
| | *Culture* | *Goals* | *Infrastructures* | *People* | *Processes* | *Technology* |
| *1. Formalize the development, integration, and use of models* | • Risk aversion<br>• Resistance to change<br>• Lack of trust in external stakeholders<br>• Lack of leadership for change | • Lack of clear vision/strategy<br>• Unstable or misaligned decision-making | • Inadequate IT infrastructure<br>• Lack of technological readiness<br>• Lack of IT support | • Lack of process knowledge<br>• Fear of loss of control<br>• Insufficient training<br>• Lack of communication & coordination<br>• Hiring and retaining qualified workforce | • Lack of standards<br>• Data quality & management issues<br>• Process interoperability issues<br>• Long time period between implementation and results | • Dependence on other technologies<br>• Usability issues<br>• Technological disruptions<br>• Resolving intellectual property issues<br>• Lack of data ownership |
| *2. Provide an authoritative source of truth* | • Bureaucratic culture<br>• Lack of commitment<br>• Silo thinking | • Focus on reducing cost<br>• Lack of institutional policy | • Lack of financial resources<br>• Inadequate IT infrastructure | • Lack of digital literacy<br>• Lack of communication & coordination<br>• Lack of process knowledge | • Data quality & management issues<br>• Process interoperability issues<br>• Lack of standards<br>• Poor estimation of total costs | • Isolated systems<br>• Security & privacy issues<br>• Dependence on other technologies<br>• Resolving intellectual property issues<br>• Lack of data ownership |
| *3. Incorporate technological innovation to improve the engineering practice* | • Resistance to change<br>• Unwillingness for upfront investment<br>• Lack of leadership for change<br>• Risk aversion<br>• Lack of innovative spirit | • Uncertainty about ROI | • Lack of financial resources<br>• Lack of technological readiness | • Fear of using technology<br>• Insufficient training<br>• Lack of digital literacy | • Restrictive laws limiting process flexibility<br>• Process complexity<br>• Lack of standards<br>• Poor estimation of total costs<br>• Long time period between implementation and results | • Technological disruptions<br>• Usability issues<br>• Dependence on other technologies<br>• Resolving intellectual property issues |
| *4. Establish infrastructure and environments* | • Silo thinking<br>• Lack of managerial support<br>• Bureaucratic culture | • Lack of institutional policy<br>• Lack of clear vision/strategy | • Inadequate IT infrastructure<br>• Lack of IT support<br>• Lack of time for transformation activities<br>• Lack of technological readiness | • Lack of communication & coordination<br>• Lack of personnel<br>• Lack of process knowledge<br>• Obtaining workforce clearance | • Process interoperability issues<br>• Data quality & management issues<br>• Lack of standards<br>• Poor estimation of total costs | • Isolated systems<br>• Security & privacy issues<br>• Lack of data ownership |
| *5. Transform culture and workforce* | • Resistance to change<br>• Organizational inertia<br>• Lack of leadership for change<br>• Silo thinking | • Lack of change management<br>• Uncertainty about ROI<br>• Lack of clear vision/strategy | • Lack of financial resources<br>• Lack of technological readiness<br>• Lack of IT support | • Lack of digital literacy<br>• Fear of job loss<br>• Fear of loss of control<br>• Lack of communication & coordination<br>• Insufficient training<br>• Hiring and retaining qualified workforce<br>• Obtaining workforce clearance | • Lack of standards<br>• Legal compliance issues<br>• Process complexity<br>• Long time period between implementation and results | • Usability issues<br>• Technological disruptions |

Below, we discuss how these barriers might deter achieving each of DoD's DE policy goals.

### 4.2.1 | Implications for Goal #1: Formalize the Development, Integration,

**and Use of Models to Inform Enterprise and Program Decision-making**
Achieving this goal requires SE organizations to embed model-based practices throughout the system lifecycle, ranging from early concept development to sustainment and eventual disposal, so that models become integral to engineering analysis and enterprise-level decisions. However, people-related barriers such as insufficient training [61,68,73], lack of process knowledge [60,62], and fear of loss of control [60,68] hinder widespread model adoption. These barriers reduce the workforce's ability to develop, validate, and maintain authoritative models, particularly when collaboration across disciplinary boundaries is required. The inability to hire and retain qualified personnel [74] further exacerbates skill gaps, especially in the aerospace and defense industries that require security clearances.

Technological and process-related factors add additional friction in the achievement of this goal. Usability issues [61], technological disruptions [65], and weak IT support [65] reduce trust in digital tools, discouraging engineers from replacing legacy methods. Moreover, dependence on diverse and disconnected technologies introduces integration complexity [60], often amplified by unresolved IP concerns [74]. From a process standpoint, interoperability issues [61,68], inconsistent standards [60,75], and poor data management [68,71] lead to inconsistencies in model fidelity, version control, and traceability. Cultural resistance [60–64], such as risk aversion [60,61,67], lack of trust in external stakeholders [69], and inadequate leadership support [61,62], creates inertia against transitioning to model-centric workflows. These technical and sociocultural impediments are often compounded by infrastructure deficits and a lack of strategic clarity, ultimately stalling progress toward formalizing model use in enterprise-level decision-making. This is more of a concern given the need for numerous organizations working together for extended periods of time for the successful management of a program or a portfolio.

#### 4.2.2 | Implications for Goal #2: Provide an Enduring, Authoritative Source of Truth

Delivering an enduring ASOT requires seamless integration of data across engineering phases and organizational silos, supported by reliable access, version control, and governance. Given the vast lifecycle of many complex systems created and maintained by the DoD, which may exceed seventy years, such as in the case of the B-52 Stratofortress or the Minuteman Missiles, this goal is particularly challenging. Furthermore, achieving this objective is undermined by human and organizational barriers such as poor communication [61,62], limited digital literacy [60,61,65], and inconsistent process knowledge [60,62], which introduce ambiguity in how data is created, interpreted, and maintained. These issues lead to divergence in model usage and reduce confidence in data integrity.

Technical and infrastructural challenges further threaten the viability of a unified ASOT.

Isolated systems [65,70], lack of IT support [65], and absence of data ownership [78] create fragmented digital environments, making it difficult to consolidate information into a single, trusted repository. Moreover, for an ASOT to remain authoritative, it must be dynamically updated as new data becomes available, necessitating ongoing data synthesis and refreshing capabilities that current infrastructure often lacks. Security and privacy concerns [65,68] also restrict data sharing, especially in mission-critical and multi-contractor environments. From a SE perspective, the lack of data standards [60,75] severely complicates synthesis across models and tools, undermining the ability to maintain consistency, traceability, and interoperability across engineering phases.

Poor data quality further compounds the difficulty of maintaining a consistent ASOT over time. Separately, poor cost estimation approaches [74] introduce a different set of uncertainties, complicating long-term planning and sustainability assessments for ASOT initiatives. These process challenges are mirrored by entrenched cultural patterns, including silo thinking [64,65], bureaucratic inertia [62,65,70], and low organizational commitment [29,55], that limit cross-functional collaboration. Without adequate infrastructure investments, data standards, and contractual mechanisms that clarify data responsibilities across the multi-organizational SE ecosystem, the establishment of a trusted ASOT remains a difficult-to-achieve aspiration.

#### 4.2.3 | Implications for Goal #3: Incorporate Technological Innovation to Improve the Engineering Practice

This goal aims to promote the adoption of emerging technologies, such as AI-enabled design and management tools, simulation platforms, and digital threads, to modernize engineering practice. However, adoption of new engineering technologies is constrained by a variety of sociotechnical barriers. People-related issues such as fear of using technology [62], insufficient training [61,68,73], and lack of digital literacy [60,61,65] inhibit the workforce's willingness and ability to engage with new tools. These anxieties often stem from previous failures or unclear communication about the value and purpose of innovation.

Technological barriers are equally significant. Usability problems [61], dependency on legacy systems [60], and ongoing disruptions [65] introduce frustration and resistance among users. Legal and process-related barriers, such as restrictive laws [60,76], process complexity [61,62,70], and lack of interoperability [61,68], add overhead to system development and integration in DE. These issues are intensified in culturally conservative organizations, where risk aversion [60,61,67], lack of change leadership [65,66], and reluctance to make upfront investments [68,69] prevail. The absence of innovation-supporting infrastructure [65,71,72] and financial flexibility [29,60] further undermines experimentation. These issues are compounded by the lack of an innovative spirit [70] and uncertainty about ROI [29,65,69], which stifles exploration of new technologies and disincentivizes long-term investment in engineering modernization. Together, these intertwined barriers can significantly hamper the integration of technological innovation into everyday practice.

### 4.2.4 | Implications for Goal #4: Establish a Supporting Infrastructure and Environments to Perform Activities, Collaborate, and Communicate across Stakeholders

Establishing robust infrastructure and collaborative environments is essential for enabling distributed teams to engage in real-time engineering activities. However, the realization of this goal is obstructed by multiple sociotechnical frictions. At the people level, insufficient personnel [62], lack of communication [61,62], and process knowledge gaps [60,62] constrain the ability to use and maintain shared platforms effectively. Workforce clearance requirements [74] can further delay onboarding of critical personnel, especially in secure environments such as Sensitive Compartmented Information Facilities (SCIFs) that are common in the defense and intelligence industry.

Technology- and process-level challenges, such as isolated toolchains [65,70], limited IT support [65], and fragmented data repositories [68,71], undermine the reliability of collaborative systems central to MBSE and other SE workflows. When the tools used for version control, model integration, and systems simulation are not interoperable or are poorly supported, it becomes difficult for distributed teams to co-develop and validate system designs effectively. Moreover, poor data quality [68,71], lack of shared standards for digital artifacts [60,75], and hidden costs in tool integration efforts further erode trust in collaborative SE platforms.

Cultural and organizational barriers, including siloed thinking [64,65] and inadequate managerial support [61,62], limit horizontal communication and restrict cross-functional knowledge exchange critical for systems-level thinking. These organizational norms are often reinforced by outdated or under-resourced IT infrastructure [65,71,72] and time constraints [60,73], leaving little room for experimentation or collaborative engagement. Such limitations disproportionately affect SE initiatives that rely on iterative feedback loops and multi-stakeholder coordination.

Finally, the absence of coherent institutional policies [65] and strategic alignment across stakeholder groups [29,60,65] contributes to duplicated modeling efforts, inconsistent use of engineering tools, and overall inefficiencies in system development lifecycles. In SE contexts, where lifecycle traceability and integration are key, these fragmented environments inhibit the realization of an enterprise-wide DE ecosystem.

### 4.2.5 | Goal #5: Transform the Culture and Workforce to Adopt and Support Digital Engineering across the Lifecycle

This goal aims to shift both workforce capabilities and organizational mindset toward one that embraces DE principles across the system lifecycle. However, this cultural transformation is challenged by deeply ingrained fears and structural barriers. Fear of job loss [60] and loss of control [60,68], compounded by insufficient training [61,68,73] and digital illiteracy [60,61,65], creates resistance among employees who view DE as a threat rather than an opportunity. Lack of communication [61,62] and difficulty recruiting qualified personnel [74] further limit workforce alignment with DE values. Ultimately, achieving this goal over the long term will require not only organizational efforts but also a fundamental shift in engineering education to produce professionals inherently equipped for a DE paradigm.

From a technological standpoint, usability challenges [61], weak IT support [65], and system unreliability [65] reduce user confidence and discourage engagement. Legal and procedural hurdles, including long implementation timelines [77], process complexity [61,62,70], and the absence of useful standards [60,75], can delay workforce transformation efforts and create the perception that progress is slow and uncertain, potentially undermining organizational commitment and external stakeholder confidence. Risk-averse attitudes [60,61,67], bureaucratic inertia [62,65,70], and lack of leadership for change [65,66] create environments that are fundamentally misaligned with the agility and experimentation required by DE. Additionally, limited financial resources [29,60] and inadequate infrastructure [65,71,72] constrain capacity-building initiatives. Uncertainty about ROI [29,65,69] and the absence of structured change management [69,70] reduce motivation for long-term behavioral shifts. To overcome these barriers, organizations must make deliberate and sustained investments – financial, strategic, and political – into systemic workforce development and cultural realignment. Without such commitment, DE transformation will remain limited in both scope and depth.

## 5 | Discussion

DE transformation is accelerating in the government and the industry with the purported benefits of enabling cheaper, faster, and better engineered systems, along with more efficient lifecycle management. Nevertheless, the majority of DT efforts are documented to result in failures [12–14], and organizations disproportionally prioritize tools and technologies to facilitate DE [27]. To shed light on the interdependent and sociotechnical nature of barriers to DE transformation, this study investigated the broader literature and provided a categorization in the dimensions of people, technology, processes, culture, infrastructure, and goals. We then mapped these barriers against DoD's Digital Engineering Strategy [7] and discussed how they relate to each of DoD's stated DE policy goals.

### 5.1 | Key Insights

Our findings highlight the nuanced nature of these barriers and call for an integrated approach that addresses both technical and social factors, as well as their interdependencies. Several key insights emerge from this study.

First, overwhelming evidence from the literature suggests that technological advancements alone are insufficient to achieve successful and sustainable DE transformation – social factors that originate from the workforce and the organization, such as workforce readiness, leadership support, and cultural alignment, are common themes that are cited for DT failures. This aligns with previous observations [13,14], which found that non-technical factors often outweigh technological ones in transformation efforts. Additionally, this supports DoD's recent actions that pursue DE competency evaluation frameworks [81], workforce training programs (i.e., Defense Acquisition University), and policy initiatives [7,8]. Nevertheless, the size and diversity of the defense ecosystem, which comprises many individual organizations with unique roles, capabilities, and contracting relationships, suggest that top-level support must be complemented by targeted efforts within these constituent entities. Individual organizations, including both prime and subcontractors, need tailored workforce development strategies that reflect their specific technical domains, maturity levels, and operational contexts. However, many of these challenges originate because engineers currently entering the workforce are not broadly trained for DE. Most educational programs remain rooted in Industry 2.0 paradigms, with DE largely confined to SE, a small minority of curricula. For long-term success, engineering education must embrace the digital paradigm, producing specialists who are equipped from the outset to operate effectively within complex, digitally enabled systems and understand their role in the larger DE ecosystem.

Second, our mapping exercise revealed that the influence of sociotechnical barriers on the DoD's DE policy goals is often diffuse and difficult to trace. For instance, while some barriers, like limited digital literacy, clearly impede workforce development, others, such as infrastructure limitations or organizational inertia, affect multiple goals simultaneously in less direct but possibly equally consequential ways. This ambiguity makes it difficult for managers and decision-makers to strategize and implement targeted interventions. Without a clearer understanding of how each sociotechnical barrier constrains progress toward specific goals, organizations risk adopting fragmented or misaligned implementation efforts. This concern is particularly acute in contexts where budgets do not directly correspond to specific policy goals, limiting managers' ability to prioritize resources effectively. Future work should aim to develop more diagnostic frameworks that trace these influences explicitly, offering a more actionable roadmap for aligning sociotechnical conditions with policy objectives.

Third, the interdependence between sociotechnical dimensions requires a holistic approach, with attention to both human-centric and technological factors. A critical area where this interdependence becomes particularly significant is tool interoperability and standardization. Many organizations face challenges when integrating diverse engineering tools across disciplines, subject matter expertise, and lifecycle phases, which can lead to inefficiencies, duplicated effort, and errors. Practical examples of these interoperability challenges underscore the complex reality of DE integration. For instance, a new DE design tool may produce output files incompatible with legacy simulation software, necessitating manual conversion or custom scripting – processes that introduce delays and risk of errors. More critically, these barriers reduce the workforce's ability to develop, validate, and maintain authoritative models. This is particularly true when collaboration across disciplinary boundaries is required, as it's typical in complex systems development. For example, a CAD model for physical components may not integrate cleanly with an MBSE model describing logical system architecture, or with a mission simulation model capturing operational behavior. Many organizations are currently struggling with these types of tool integration efforts. Without seamless data exchange and compatible toolchains, teams face significant friction in constructing end-to-end digital representations, undermining one of DE's core promises. Addressing these challenges requires explicit attention to SE practices, such as defining interface standards, establishing model verification and validation workflows, and ensuring that tools support cross-disciplinary collaboration.

Fourth, the successful implementation of DE relies on strategic leadership and sustained practices that promote a culture of innovation, a more risk-neutral approach, and methods and processes to support interdisciplinary collaboration. It also requires patience in the face of uncertain or long-term returns on investment. Given the emergent and evolving nature of DE practices, organizations must be willing to commit to iterative development and learning, even when the benefits may not be immediately quantifiable. Organizational inertia, resistance to change, inadequate IT infrastructure, and expectations for immediate benefits emerged as critical impediments. These are difficult barriers to navigate, particularly given that policymakers and organizational leaders often serve limited terms in leadership roles. This situation is especially pronounced in government agencies, where leaders are typically appointed on rotational terms; for example, a Technical Director at a Navy Engineering Agency or a Pentagon official may serve less than 5 years in a given role. Knowing they will not be in office long, these short appointments may lead to a prioritization of shorter-term wins over long-term DE initiatives, further exacerbating challenges in sustaining organizational change. This brings forth an inherent risk where each new administrator might seek new implementation practices and discontinue previously implemented initiatives. This creates pressure to achieve short-term wins, particularly when long-term DE efforts require substantial investment from their own budgets, further complicating sustained organizational change. Such pressures are generally less pronounced in industry, where leadership continuity tends to be greater.

Fifth, while this study did not systematically analyze the interplay between individual barriers, several cross-dimensional interactions stood out during the review process and merit further attention. For example, people-related barriers such as limited digital literacy and insufficient training often amplify technology-related issues like poor usability and unstable tool performance. These interactions can create feedback loops where users disengage from DE environments not solely because of the tools themselves, but due to a combination of inadequate training, fear of job loss, and weak IT support. Similarly, infrastructure-related constraints, such as poor system responsiveness or lack of connectivity, can entrench siloed thinking and cultural resistance, especially in organizations with a history of failed IT rollouts. These patterns illustrate how barriers reinforce one another across dimensions and underscore the need for DE implementation strategies that recognize and address these cascading effects. While we recognize the importance of these cross-dimensional interactions, a comprehensive analysis of their full scope is beyond the focus of this study.

## 5.2 | Implications

### 5.2.1 | Implications for Engineers, Managers, and Policymakers

For *engineers* working "in the trenches," this study reinforces the fact that the majority of the barriers to DE transformation do not stem from individual shortcomings but from systemic organizational, technical, and cultural misalignments. However, engineers should recognize that embracing DE can ultimately benefit their daily work, providing opportunities to improve their workflows, enhance collaboration, and contribute to more efficient and effective system development. Thus, by actively engaging with the transformation efforts and acting as good "citizens" of the new DE ecosystem, they can play a meaningful role in realizing these benefits. Practicing SEs should be encouraged to view DE not as a threat but as an evolving practice that, if well-implemented, can reduce rework, improve collaboration, and increase visibility into system behavior. Understanding the structural roots of implementation challenges may help reduce resistance, build trust in new tools, and empower engineers to shape the transformation from within.

For organizational *managers*, the sociotechnical framework and categorization of barriers provided in this study present a structured lens to anticipate and mitigate risks in DE implementation. The six-dimensional categorization of barriers could be utilized as a "playbook" to diagnose pain points and design more resilient transformation strategies. The mapping of these barriers to the DoD's DE goals enables targeted, goal-oriented interventions that can help organizations better allocate resources, align internal strategies, and evaluate implementation progress [7,9]. This can also serve as a practical tool for negotiating budgets and for communicating both challenges and priorities with policymakers and engineers on the ground. This guidance is particularly relevant for managers tasked with integrating DE in mission-driven or safety-critical domains, where lifecycle complexity and interdependence demand sustained organizational learning and cross-functional integration.

For *policymakers*, this study underscores that DE transformation cannot be achieved by policy mandates alone. While the DoD and related agencies have made significant progress in setting expectations through guidance documents and implementation strategies, one of the central findings of this study is that these efforts often lack the long-term financial commitment and programmatic support required for lasting transformation. This observation echoes a broader concern that without addressing these multifaceted barriers, the infrastructure needed to support enterprise-wide DE transformation cannot be fully realized; and without implementing structural enablers, even well-designed DE initiatives risk stagnation or reversal. In that regard, the approach developed here provides empirical backing for these concerns, highlighting financial, human, and infrastructural constraints as recurring themes in documented DE failures. Moreover, our work identifies the mechanisms through which policymakers can support managers in overcoming these barriers and offers a structured approach for evaluating managerial performance in implementing DE initiatives.

#### 5.2.2 | Implications for Researchers

For researchers in the fields of SE and DE, this study addresses several pressing gaps in the literature. Prior research has noted that the benefits of MBSE, and by extension DE, are often based more on perception than empirical evidence, with limited studies providing measurable or observed support [25,26]. By identifying sociotechnical variables that influence DE implementation across organizational contexts, our findings offer a direct response to this concern. As a key contribution, this study provides a structured and testable set of barriers, offering researchers a concrete, albeit non-exhaustive, foundation for understanding the factors that shape DE transformation. In addition, by translating these barriers into a coherent framework, this work contributes to the operationalization and maturation of the DE research agenda, enabling more directed empirical investigation and the development of measurement approaches to assess their impact.

Calls for systematic measurement frameworks that can link DE initiatives to meaningful outcomes have also gained traction in recent studies [10,82]. Such frameworks are critical for moving beyond anecdotal or perception-based assessments, enabling organizations to evaluate the effectiveness of DE practices in achieving strategic and operational objectives. By identifying and contextualizing the organizational and cultural barriers that influence DE implementation, our study provides a foundation for designing metrics that are both empirically grounded and practically actionable. The accompanying policy mapping further extends this contribution by providing a mechanism to examine how strategic objectives intersect with operational constraints, an area often overlooked in existing metrics-focused research.

More broadly, this study lays foundational groundwork for future empirical research on the sociotechnical dynamics of DE transformation. The structured barrier framework and policy mapping approach introduced in this study create opportunities for researchers to pursue diverse study designs, such as cross-sectoral surveys, longitudinal analyses, and case-based investigations. The findings also underscore the utility of integrative research methods that bridge technical systems thinking with insights from organizational science, policy analysis, and human factors.

As DE continues to reshape the development and management of complex engineered systems, researchers play a critical role in advancing evidence-based strategies for implementation, policy design, and impact evaluation. Future research, grounded in a sociotechnical awareness of the SE context and informed by empirical rigor, can help close the gap between DE's transformative potential and its realized outcomes.

### 5.3 | Limitations

This study provides a structured understanding of sociotechnical barriers to DE transformation and maps these barriers against the DoD's DE goals, offering a practical playbook. However, we acknowledge two main limitations. *First,* our analysis is based on a rapid review of existing literature, which, although effective for synthesizing insights promptly, may have missed some relevant studies due to its scope and reliance on *Google Scholar* as the sole database compared to a comprehensive systematic review. Nevertheless, given that DE is a relatively new concept, relying on *Google Scholar* may not be an overly restrictive approach at this stage. *Second,* our research approach draws from secondary data sources across diverse industries. Although this is an intentional generalizability choice, it may have introduced context-specific nuances that are not entirely representative of DE efforts in organizations that are engaged with the development and management of complex engineered systems, particularly in highly regulated sectors like defense or aerospace. While we tried to bridge this gap in our discussion, our findings should be taken with a grain of salt, and there could be some deviations. Future empirical research will need to verify the extent to which these barriers hold for the SE community.

### 5.4 | Future Work

Future research could pursue several directions to advance the field. One promising avenue is the empirical validation of the identified DE barriers. A comprehensive, cross-sector survey of SE practitioners in industries such as defense, aerospace, energy, and transportation could reveal how these sociotechnical challenges manifest in practice and assess the relative significance of each barrier across organizational contexts. This kind of study could also enable a deeper investigation into the workforce, particularly how varying levels of expertise and experience shape practitioners' perceptions of DE challenges. Such empirical grounding would enhance the practical relevance and applicability of the framework. In addition, longitudinal studies could provide insights into how sociotechnical barriers emerge, evolve, and interact over time across the DE lifecycle. Since transformation is not a static event but a process unfolding across design, implementation, and sustainment phases, longitudinal research could clarify which barriers are transient and which persist, and how their trajectories are shaped by organizational decisions or external pressures.

Another potential direction involves investigating practical *tactics* for overcoming DE barriers. This includes examining organizations that have made measurable progress in their DE journeys to extract actionable lessons about leadership practices, change management tactics, stakeholder engagement mechanisms, and infrastructure investments. Identifying what works in overcoming sociotechnical resistance will be key to enabling sustainable and resilient transformation. Additionally, case studies of private-sector organizations, where DT efforts have often been more mature, could yield valuable insights into effective practices, though accessing proprietary data remains a methodological challenge. Finally, future research could extend the sociotechnical barrier framework and policy mapping approach beyond the DoD. Mission-driven federal agencies such as the Department of Energy (DOE), NASA, and the Department of Homeland Security face similar transformation pressures but operate under different mandates, cultures, and technical environments. Applying and adapting the framework in these settings could reveal new classes of barriers or policy gaps, thus broadening the framework's utility and strengthening its generalizability.

## 6 | Conclusion

This study advances the understanding of DE transformation by systematically identifying and categorizing sociotechnical barriers across six dimensions – people, technology, processes, culture, infrastructure, and goals – and mapping them to the DoD's DE policy goals. Through this categorization and mapping, we offer a structured view of where DE transformation challenges could arise and how they may undermine strategic priorities. Our findings reaffirm that DE transformation is not merely a technical challenge but a deeply sociotechnical endeavor, where entrenched organizational cultures, skill gaps, fragmented goals, and tool interoperability issues play central roles. Drawing from both DE-specific and broader DT literature, our study bridges conceptual insights with operational relevance, offering a more integrative foundation for diagnosing DE transformation challenges in complex engineering environments. By recognizing the multifaceted nature of these barriers and their impact on DE goals, organizations can better navigate the complexities of transformation and create environments conducive to innovation, collaboration, and technological advancement [8,83].

The sociotechnical barrier categorization and policy mapping developed here serve as a diagnostic lens for identifying where systemic misalignments occur and how they propagate across organizational levels. While the focus of this work has been the DoD, the framework and findings are likely transferable to other mission-driven agencies and sectors undertaking similar transformations. As organizations continue to adopt DE in pursuit of more agile, integrated, and model-driven engineering practices, it is imperative to recognize that transformation success hinges not solely on technological upgrades but on aligning people, structures, and strategies.